\documentclass[aps,prl,twocolumn,longbibliography]{revtex4-1}
\usepackage{natbib}
\usepackage{graphicx,color,xcolor,dcolumn,bm,amsfonts,amsmath,amssymb}
\usepackage{mathptmx,longtable,float,soul}
\usepackage{titlesec,booktabs,eqnarray}
\usepackage{mathptmx,longtable,float}
\usepackage{multirow,soul}
\usepackage{xr}
\definecolor{orange}{rgb}{1,0.5,0}
\definecolor{mypink}{rgb}{0.858, 0.188, 0.478}
\definecolor{mygray}{gray}{0.6}

\begin{document}
\title{Microscopic Origin of Polarization-Controlled Magnetization Switching in FePt/BaTiO$_3$}
\author{Qurat ul ain$^{1,2}$}
\email[Email address: ]{quratulain@qau.edu.pk}
\author{Thi H. Ho$^{3}$}
\author{Soon Cheol Hong$^{2}$}
\author{Dorj Odkhuu$^{4}$}
\author{S. H. Rhim$^{2}$}
\email[Email address: ]{sonny@ulsan.ac.kr}
\affiliation{
$^{1}$ Department of Physics, Quaid-i-Azam University, Islamabad, 45320, Pakistan\\
$^{2}$Department of Physics and Energy Harvest-Storage Research Center, University of Ulsan, 
Ulsan, Republic of Korea\\
$^{3}$Laboratory for Computational Physics, Institute for Computational Science and Artificial Intelligence, Van Lang University, Ho Chi Minh City, Vietnam\\
$^{4}$Department of Physics, Incheon National University, Incheon, Republic of Korea\\
}
\date{\today}

\begin{abstract}
Electric-field driven magnetization switching in FePt/BaTiO$_3$ (001) is demonstrated through first-principles calculations.
The magnetic easy axis of FePt layer undergoes a transition from in-plane to perpendicular direction 
upon ferroelectric polarization reversal, a process sensitively controlled by epitaxial strain 
with threshold strain strain($\eta$) $\eta\approx\%$. 
At this phenomena, 
a large interfacial magnetoelectric coupling ($\alpha_I = 3.6 \times 10^{-10}$ G$\cdot$cm$^2$/V) is responsible, 
stemming from the orbital reconstruction.
In particular, the redistribution of Pt-$d$ orbital occupancy alters spin-orbit coupling, 
thereby tuning the competition between magnetic anisotropy ($K_i$) and magnetoelastic energy ($b_1$). 
Our work clarifies the fundamental physics of strain-engineered magnetoelectricity and suggests a concrete pathway for designing ultra-low-power voltage-controlled magnetic memory.
\end{abstract}
\maketitle

\section{Introduction}
\label{sec:intro}
The quest for "Beyond CMOS technology" has intensified the search for voltage-controlled magnetism and energy-efficient material platforms for next-generation information processing\cite{theis2010s,nikonov2015benchmarking,manipatruni2018beyond}. 
Among these concepts, the magnetoelectric spin-orbit (MESO) device\cite{manipatruni2019scalable} 
combines magnetoelectric (ME) switching for electric-field control with spin-orbit-based readout that exploits inverse Rashba-Edelstein and inverse spin Hall effects\cite{edelstein1990spin,sanchez2013spin,omori2014inverse}.
To realize such low-power operation, a robust coupling between electric polarization and magnetization is essential, particularly within thin-film architectures compatible with device integration.

Composite multiferroic heterostructures, comprising ferroelectric (FE) and ferromagnetic (FM) stacks, 
provide a practical platform for electric-field control of magnetism\cite{eerenstein2006multiferroic,zheng2004multiferroic}. 
Among various magnetoelectric (ME) coupling channels, 
the strain-mediated route is notably robust and scalable\cite{sheoran2019comparative,cheng2018recent,shvartsman2011converse}. 
In this approach, electric-field-induced strain in the FE layer is transmitted across the interface to the FM, modulating its magnetic anisotropy through magnetoelasticity\cite{wang2010electric,hu2009electric,hu2015purely}. 
This mechanism is most effective when an FE layer with a large piezoelectric response is coupled to an FM with strong magnetostriction.

FePt/BaTiO$_3$ (BTO) is an attractive platform in this regard: ferroelectric BTO exhibits large piezoelectricity\cite{haeni2004room,choi2004enhancement}, 
while FePt possesses large saturation magnetization, strong magnetocrystalline anisotropy, and appreciable magnetostriction\cite{Okamoto2002,thiele1998perpendicular,spada2003}. 
While voltage-controlled magnetization reorientation has been already demonstrated\cite{yang2017surface, leiva2022electric}, 
the underlying microscopic mechanism deserves investigation. 
In particular, the competetion between 
between magnetic anisotropy ($K_i$) and magnetoelastic energy ($b_1$) has not been established.

In this work, first-principles electronic-structure calculations are employed to uncover strain-driven polarization control of magnetization in FePt/BaTiO$_3$(001). 
We predict  a sizable interfacial magnetoelectric coefficient, $\alpha_I = 3.6 \times 10^{-10}$ G$\cdot$cm$^2$/V, originating from polarization-induced orbital reconstruction at the interface. 
Crucially, polarization reversal, coupled with epitaxial strain, switches the FePt easy axis between perpendicular and in-plane. 
This switching, governed by the interplay between interfacial uniaxial anisotropy ($K_2^i$) 
and the first-order magnetoelastic coefficient ($b_1$), reveals the microscopic origin of voltage-controlled magnetization switching in FePt/BTO. These results provide fundamental design guidance for low-power spintronic devices based on strain-engineered interfaces.

\section{Computational methods}
\label{sec:comp-meth}
\begin{figure}[htbp]
\centering
   \includegraphics[width=\columnwidth]{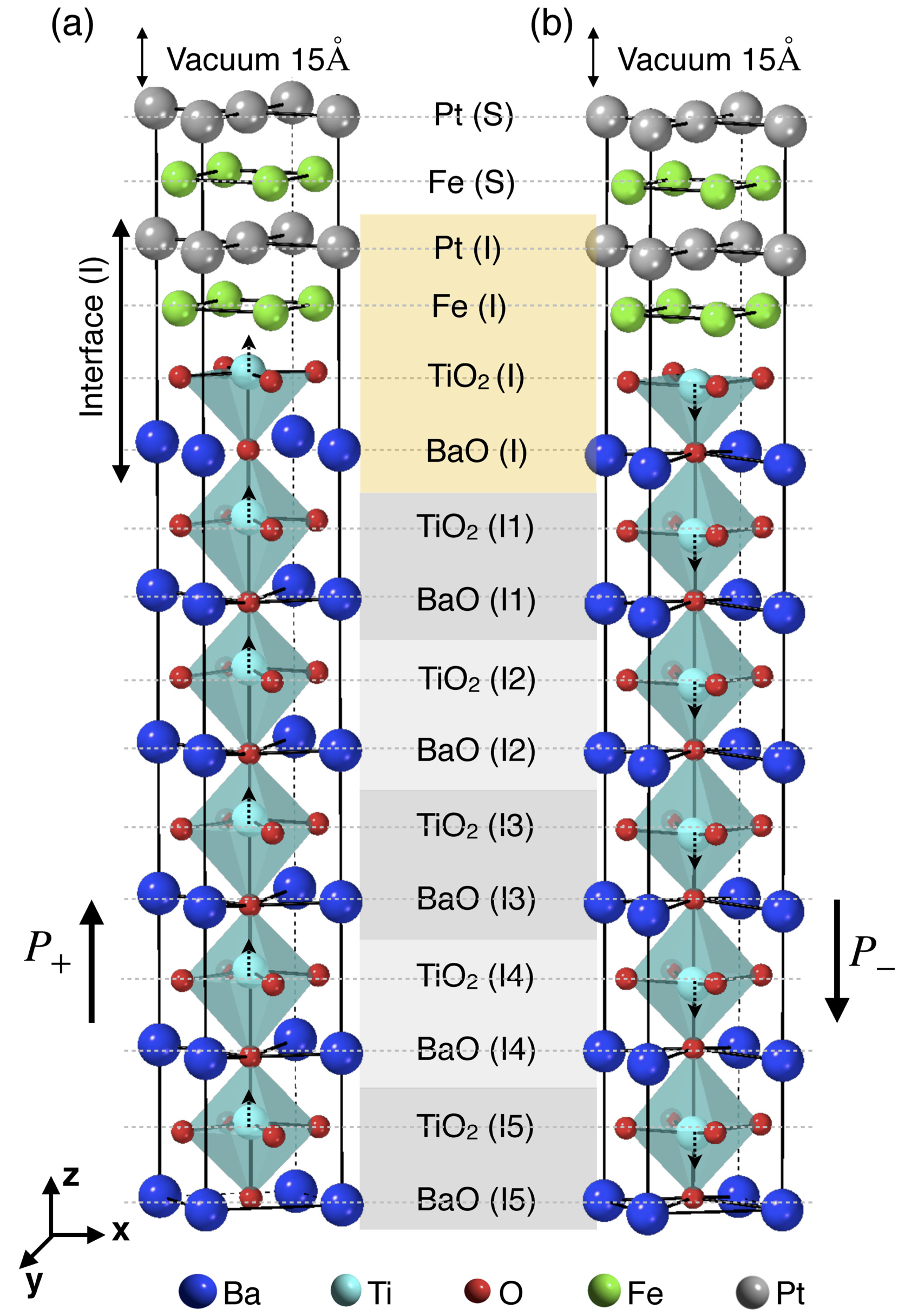}
   \caption{\label{fig01} Schematics of the FePt/BaTiO$_3$(001) heterostructure: (a) $P_+$ and (b)  $P_-$, as indicated by black solid arrows. The shaded central panel highlights the regions of the  surface (S), interface (I), and BaTiO$_3$ layer. Grey, green, blue, cyan, and red spheres represent  Pt, Fe, Ba, Ti, and O atoms, respectively. Dotted arrows within the octahedra indicate the local  displacements of atoms.}
\end{figure}

Density functional theory (DFT) calculations are performed using the the Vienna {\em ab initio} simulation package (VASP) \cite{kresse1996efficient} with the plane-wave projector augmented wave (PAW) basis \cite{blochl1994projector}.
The exchange-correlation potential is treated using the generalized gradient approximation (GGA) of Perdew, Burke, and Ernzerhof \cite{Perdew1996PBE}. Energy cutoff of 450 eV is used, with Gaussian smearing $\sigma = 0.05$ eV, and {\em k}-point mesh of 15$\times$15$\times$1 is employed for structural relaxation.

The FePt/BaTiO$_3$ heterostructure is modeled using 6 monolayers (ML) of FePt(001) on 12 ML (equivalent to 5.5 unit cells) of BaTiO$_3$ (001) with a 15 $\AA$ vacuum region. 
As illustrated in Fig. \ref{fig01}, the interface (I) region consists of single Fe, Pt, TiO$_2$, and BaO layers. 
Fe atoms occupy top site on the O atom of TiO$_2$ layer. 
All layers are allowed to relax except for four bottom-most BTO unit cells, which are fixed with their bulk polarization. The net polarization along the $z$-axis is defined as $P_+$ for $z > 0$ and $P_-$ for $z < 0$, as indicated by the solid black arrow in Fig. \ref{fig01}.

Optimized in-plane lattice constants of 3.99 $\AA$ for bulk BaTiO$_3$ and 3.87 $\AA$ for FePt are used, which agree well with experiment\cite{dungan1952lattice,sakuma1994first}. Accordingly, 3\% lattice mismatch is observed for in-plane FePt with a lattice constant of 3.87 $\AA$. To investigate the effect of external strain on the magnetic properties, strain ($\eta$) is varied from $-2\%$ to $3\%$, defined as: $\eta = (a - a_o) / a_o$, where $a_o$ is the equilibrium lattice constant of bulk BaTiO$_3$. Interlayer distances are relaxed for each strain with a force convergence criterion of 1$\times 10^{-3}$ eV/$\AA$.

Spin-orbit coupling (SOC) is included in the second-variational method \cite{koelling1977technique}. The magnetic anisotropy energy ($E_{MA}$) is estimated by $E_{MA} = [E_{100} - E_{001}] / A$, where $E_{100}$ and $E_{001}$ are the total energies with magnetization along the [100] and [001] directions, respectively \cite{koelling1977technique}; $A$ is the surface area. The magnetic anisotropy energy $E_{MA}$ includes contributions from magnetoelastic and uniaxial anisotropies. For reliable values of $E_{MA}$, a denser {\em k}-point mesh of 33$\times$33$\times$1 is used.

\section{Results and Discussion}
{\label{sec:results}
\begin{figure}[htbp]
\begin{center}
  \includegraphics[clip=true,width=\columnwidth]{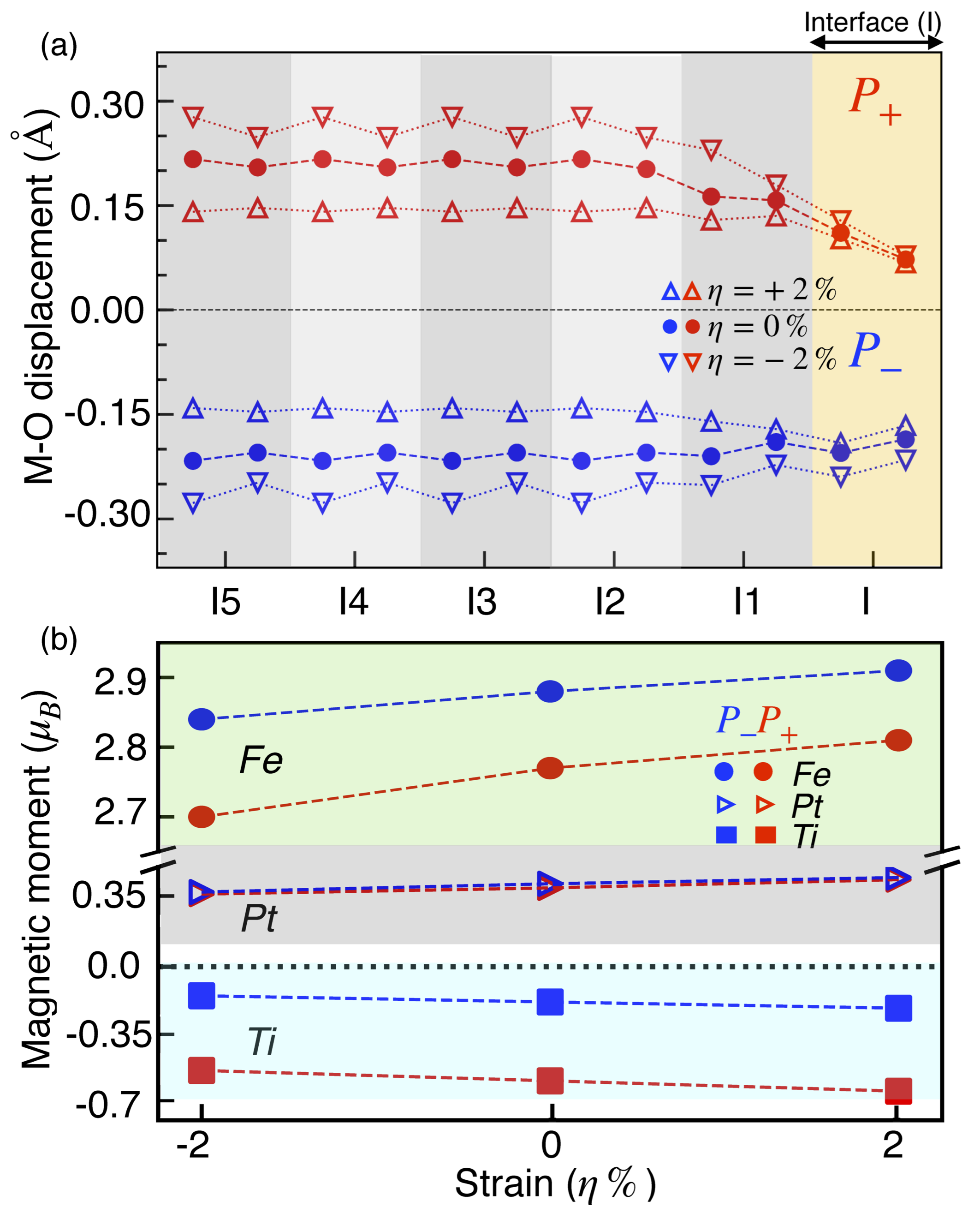}
  \caption{\label{fig02} (a) Relative displacement of cations (M = Ba and Ti) with respect to O plane. Filled circles and upper(lower) triangles are for 0$\%$ and +2(-2)$\%$ strains. Shaded region corresponds to the interface (I) and BaTiO$_3$ regions. (b) Magnetic moment ($\mu_B$) of interface Fe, Pt, and Ti  atoms as a function of strain. Broken y-axis is separate Fe from Pt and Ti with colored shades. Circle, triangle, and square represent Fe, Pt, and Ti atoms, respectively. Red and blue denote $P_+$ and $P_-$, respectively. }
\end{center}
\end{figure}

The relative displacements of cations (M = Ba and Ti) are plotted in Fig.~\ref{fig02}(a) with respect to neighboring oxygen (O) at each plane under 0\% and $\pm$2\% strain for both polarization directions. Displacements along $z > 0$ ($z < 0$) correspond to cations moving towards (away from) the interface and are represented by red and blue symbols, respectively. At the interface, the M-O displacement reduces abruptly for $P_+$ compared to the bulk, while it remains almost unaffected for $P_-$ irrespective of strain. Nonetheless, the relative displacements between cations and O atoms decrease (increase) with tensile (compressive) strain.

Magnetic moments of the interfacial Fe, Pt, and Ti atoms at different strain values are plotted in Fig.~\ref{fig02}(b), with red and blue symbols for $P_+$ and $P_-$, respectively. At zero strain, magnetic moment of the interfacial Fe is 2.77 $\mu_B$ for $P_+$ and 2.88 $\mu_B$ for $P_-$, respectively. This difference in magnetic moment results from the structural sensitivity associated with the asymmetric interface-induced electric polarization. The Ti atom induces a magnetic moment of -0.55 $\mu_B$ for $P_+$, which is significantly larger than the -0.17 $\mu_B$ for $P_-$. In contrast, Pt moment show little change under polarization reversal.

The interface magnetoelectricity is characterized by the coupling coefficient $\alpha_I = \mu_o \Delta \mu_s / \mathcal{E}$, which quantifies the difference in induced interfacial magnetization via electric polarization reversal. Here, $\mu_o$ is the vacuum permeability, $\Delta \mu_s$ is the difference in magnetic moment between $P_+$ and $P_-$, and 
$\mathcal{E}$ is the external electric field \cite{niranjan2008magnetoelectric,fujita2019first,niranjan2010electric}. We use the ferroelectric coercive field $\mathcal{E}_c = 120$ kV/cm from experiment \cite{Radaelli2014Electric}. The interface magnetoelectric coupling coefficient for FePt/BaTiO$_3$ is found to be $\alpha_I = 3.6 \times 10^{-10}$ G$\cdot$cm$^2$/V, which is much higher than the 2.1 $\times$ 10$^{-10}$ G$\cdot$cm$^2$/V for Fe/BaTiO$_3$ \cite{duan2006predicted} and 2.0 $\times$ 10$^{-10}$ G$\cdot$cm$^2$/V for Co/Pb(Zr,Ti) \cite{vlasin2016interface}.

\begin{figure}[htbp]
\begin{center}
  \includegraphics[clip=true,width=\columnwidth]{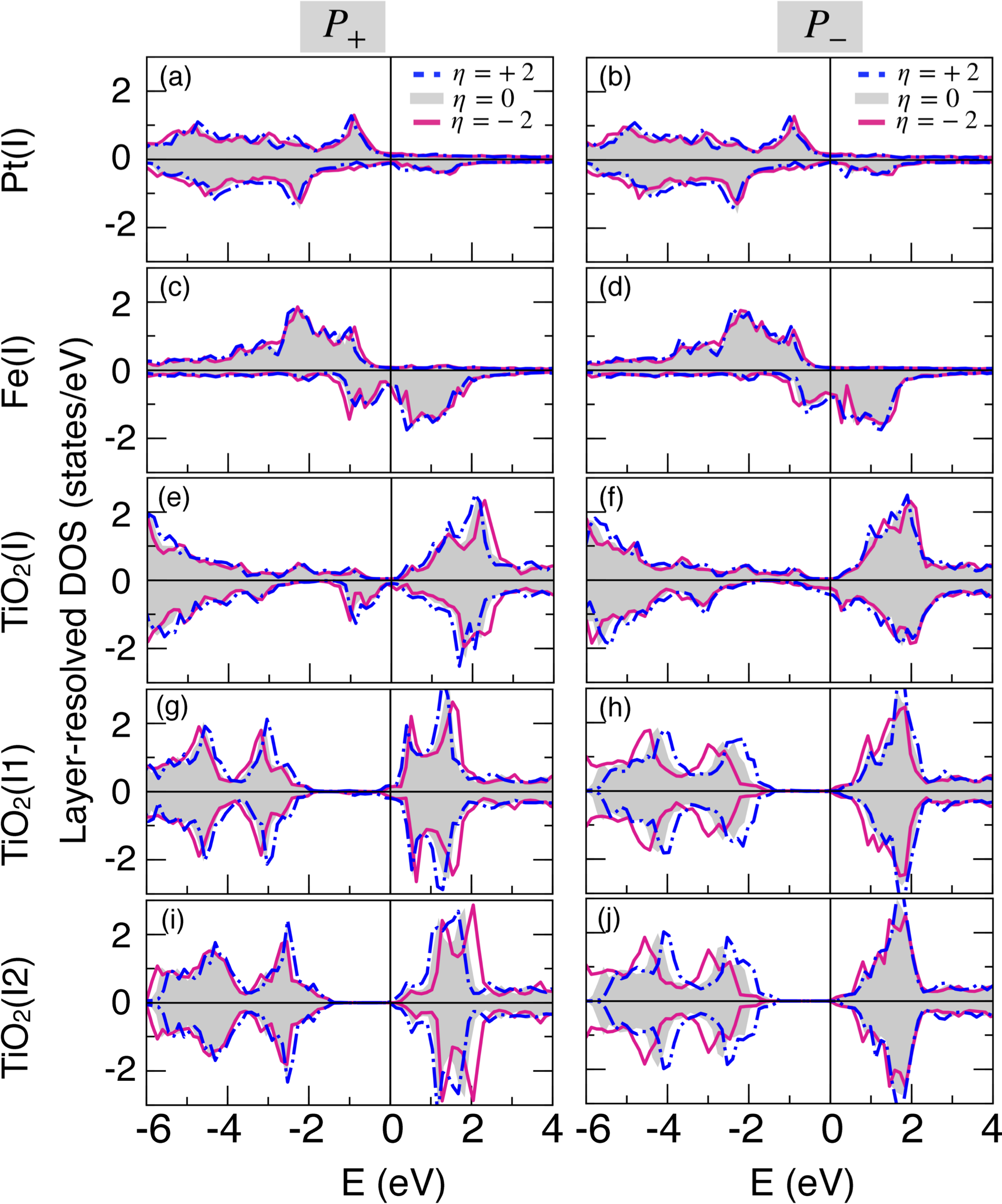}
   \caption{\label{fig03} Layer-resolved density of states (LDOS) for the interfacial (I) region: (a) and (b) Pt(I), (c) and (d) Fe(I), (e) and (f) TiO$_2$(I), and for a few inner layers: (g) and (h) TiO$_2$(I1), (i) and (j) TiO$_2$(I2), for $P_+$ (left panels) and $P_-$ (right panels). The filled grey region represents the LDOS at zero strain. The dashed blue and solid pink lines represent the DOS at +2$\%$ and -2$\%$ strain, respectively. The vertical solid line indicates the Fermi level ($E_F$), which is set to zero.}
\end{center}
\end{figure}
To understand the interface magnetoelectric effect, layer-resolved density of states (LDOS) are plotted in Fig.~\ref{fig03} for $P_+$ (left panel) and $P_-$ (right panel) at $\eta$ = 0 and $\pm$2\%, respectively, including LDOS of the interface layers (Pt, Fe, TiO$_2$) and inner layers (TiO$_2$(I1) and TiO$_2$(I2)). 
LDOS of BaO is not shown, as the states of BaO layers are far below the Fermi energy ($E_F$).

For Fe (I), the minority-spin states exhibit strong sensitivity to polarization reversal, while the majority-spin channel remains nearly unchanged, as reflected in LDOS near $E_F$. Specially, DOS at $E_F$ is higher for $P_-$ than for $P_+$, indicating a polarization-dependent modulation of spin states at the interface.
For TiO$_2$ layers away from the interface, a clear insulating behavior persists for both polarizations. 
In contrast, TiO$_2$(I) layers are distinct: for $P_-$, minority-spin states appear at $E_F$, suggesting weak spin-polarized metallicity localized at the interface. On the other hand, for $P_+$,  the minority-spin states of TiO$_2$ (I) are pushed  below $E_F$, resulting in nearly zero DOS at $E_F$, consistent with the insulating nature of the BaTiO$_3$. Pt (I) DOS remains largely unchanged upon polarization reversal, implying that the polarization primarily alters the interfacial electronic structure within a few atomic layers. Strong hybridization occurs between the minority spin Ti-$d_{xz/yz}$ and majority spin Fe-$d_{xy}$ states. The Ti-Fe $d$-orbital overlap is more significant for $P_+$ than for $P_-$, leading to a higher magnetic moment for $P_+$ compared to $P_-$. Furthermore, Pt atom induces finite magnetic moment due to Pt-$d_{x^2-y^2}$ orbital hybridization with Fe-$d_{xz/yz}$ as shown in Fig. S1 and S2 in Supplementary Material\cite{DOS1}. Finally, O-2$p$ orbitals lie well below $E_F$ and show only weak overlap with Fe-3$d$ states, resulting in a small induced magnetic moment of $\sim$ 0.06 $\mu_B$. The extent of hybridization depends on the strength of the orbital overlap, inversely on the energy separation between orbitals.

Calculated magnetic anisotropy energy ($E_{MA}$) is plotted in Fig.~\ref{fig04}(a) for both polarization directions as a function of strain. $E_{MA}$ varies linearly with strain. At $\eta=0\%$, $E_{MA}$  are 7.15 and 8.15 erg/cm$^2$ for $P_+$ and $P_-$, respectively. Notably, around $\eta=2\%$, when the polarization changes from $P_-$ to $P_+$ there is a sign change of $E_{MA}$, the magnetization switches from perpendicular to in-plane. This behavior is consistent with a phenomenological model \cite{qurat2021JOM}, 
which treats the total magnetic energy as a sum of magnetoelastic, magnetocrystalline, and demagnetization contributions.  In the model, the spin reorientation transition arises from the competition between these terms, 
yielding a critical strain of approximately $\text{2\%}$, in an excellent agreement with our present first-principles results.

Total magnetic anisotropy energy is  written conveniently as
\begin{equation}
\label{eq:MEL}
E_{MA} = E_{MA}^0 + b_1 t \sum_{k=1}^{3} \eta_k \alpha_k^2,
\end{equation}
where $E_{MA}^0$ is the magnetic anisotropy energy per area; $\eta_k$ ($k=1,2,3$) are the strain tensor; $\alpha_k$ the direction cosines of magnetization; $t$ is the FM film thickness; $b_1$ is the first-order magneto-elastic (MEL) coefficient\cite{principles}; the superscript 0 denotes zero strain. 
In the tetragonal structure, since $\eta_1=\eta_2$, the perpendicular strain $\eta_3$ is determined from the magneto-elastic equation of state\cite{qurat2020voltage}.
$E_{MA}^0$ includes contributions of the uniaxial anisotropy, 
$K_2^i(1-\alpha_3^2)$, and the shape anisotropy, $K_{sh}$. In the thin film limit, the interface/surface energy dominates over the bulk counterpart. The surface contribution to shape anisotropy $K_{sh}$, using Bruno's relationship\cite{bruno1993magnetismus} is:
 $K_{sh} = -\frac{1}{2} \mu_o M_s M_v$,
where $M_s$ is the sum of excess surface magnetization per unit area for each layer and $M_v$ is the bulk magnetization.Strain obtained from Eq.~(\ref{eq:MEL}) gives:
\begin{equation}
\label{eq:MA2}
E_{MA} = K_{eff} + 2 b_1 t \eta, 
\end{equation}
where $K_{eff} = K_2^i + K_{sh}$.
As shown in Fig.~\ref{fig04}(a), $K_2^i(1-\alpha_3^2)$ and ME coefficient ($b_1$) are extracted.
The uniaxial anisotropy, $K_2^i$, favors perpendicular magnetization $\sim$ 7.35 and 8.44 erg/cm$^2$ for $P_+$ and $P_-$, respectively. The calculated shape anisotropy contribution is approximately -0.25 erg/cm$^2$ for both polarizations, which is small and nearly strain-independent. 
$K_{sh}$=-0.31 erg/cm$^2$ is consistent with that of 3 ML Fe/BaTiO$_3$ film\cite{odkhuu2018strain}. 
Notably, the first-order magnetoelastic coefficient $b_1$ =0.18$\times$10$^{8}$ erg/cm$^3$ is sufficiently large to compete with the $K_2^i$ and 
thereby induce magnetization switching at $\eta = 2\%$.

\begin{figure}[htbp]
\begin{center}
   \includegraphics[width=\columnwidth]{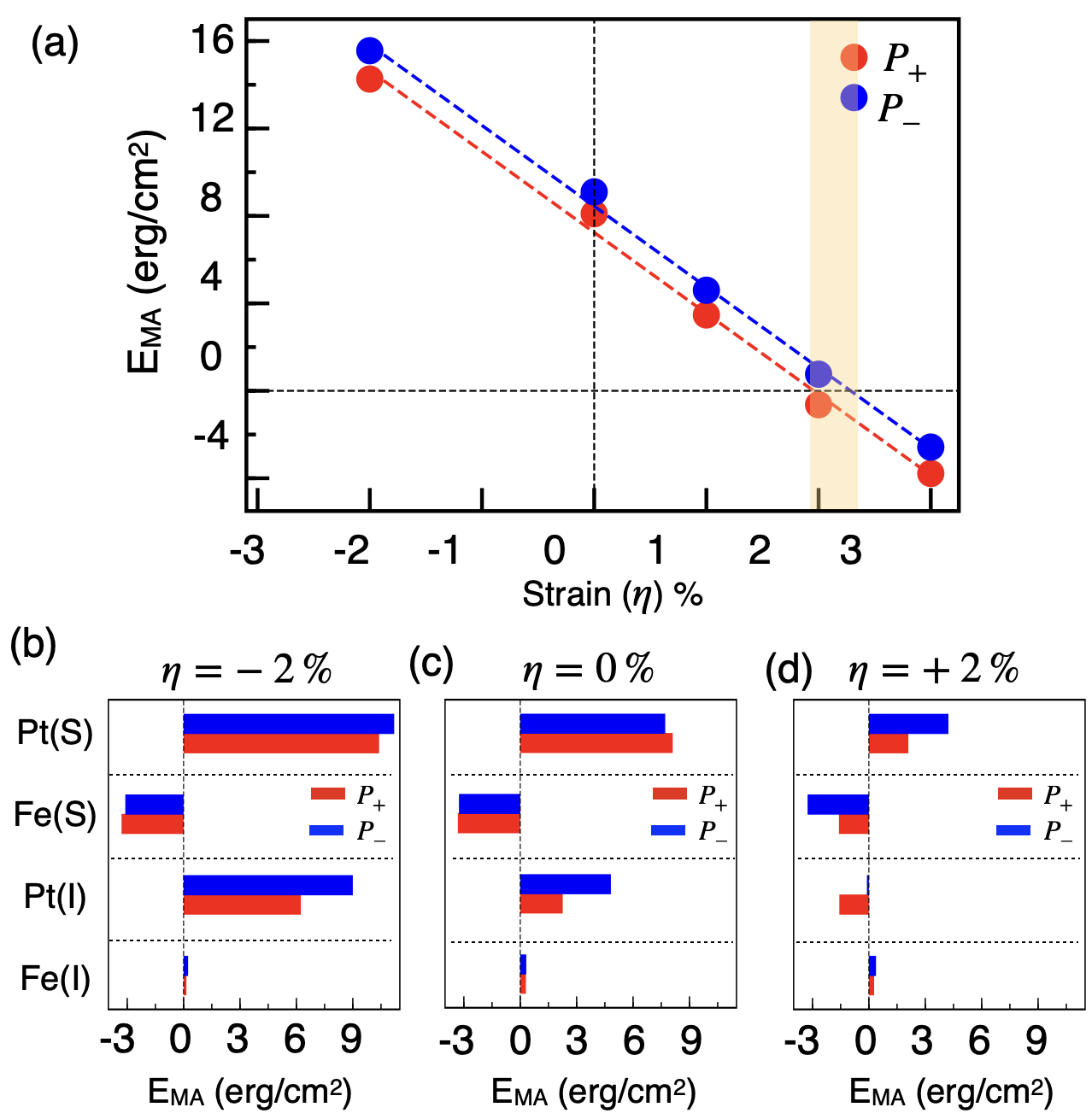}
\caption{\label{fig04}Magnetic anisotropy energy ($E_{MA}$) of FePt/BaTiO$_3$ heterostructure as a function of strain ($\eta$). 
  Red and blue color denote $E_{MA}$ for $P_+$ and $P_-$, respectively.
  Shaded region indicates critical strain ($\eta_c$) at which magnetic anisotropy energy shifts from perpendicular to in-plane direction. 
  Layer-resolved $E_{MA}$ at (b) $\eta=0\%$ and (c) $\eta=2\%$ for polarization direction $P_+$ (red) and $P_-$ (blue), respectively.}
  \end{center}
\end{figure}

The layer-resolved atomic contributions to $E_{MA}$ at $\eta$=  0$\%$ and +2$\%$ are shown in Fig.~\ref{fig04} (b-c), respectively. 
Pt atoms contribute positively to $E_{MA}$ $(>0)$ at zero strain and play a dominant role in driving the magnetization switching. 
Under tensile strain, Pt(S) contribution decreases significantly, while Pt(I) shifts to favor in-plane orientation. 
Fe atom primarily contributes to $E_{MA} (<0)$. 
Under strain, Fe(S) contribution noticeably reduces in magnitude for $P_+$, 
whereas Fe(I) contribution is negligible regardless of strain. 
The magnetization switching originates from the competition between uniaxial and magnetoelastic anisotropies, mainly driven by strain-induced modification of the interfacial Pt-Fe orbital hybridization and redistribution of Pt- $d$ states near $E_F$ as shown in Fig.~\ref{fig03} 

\begin{figure}[htbp]
\begin{center}
  \includegraphics[width=\columnwidth]{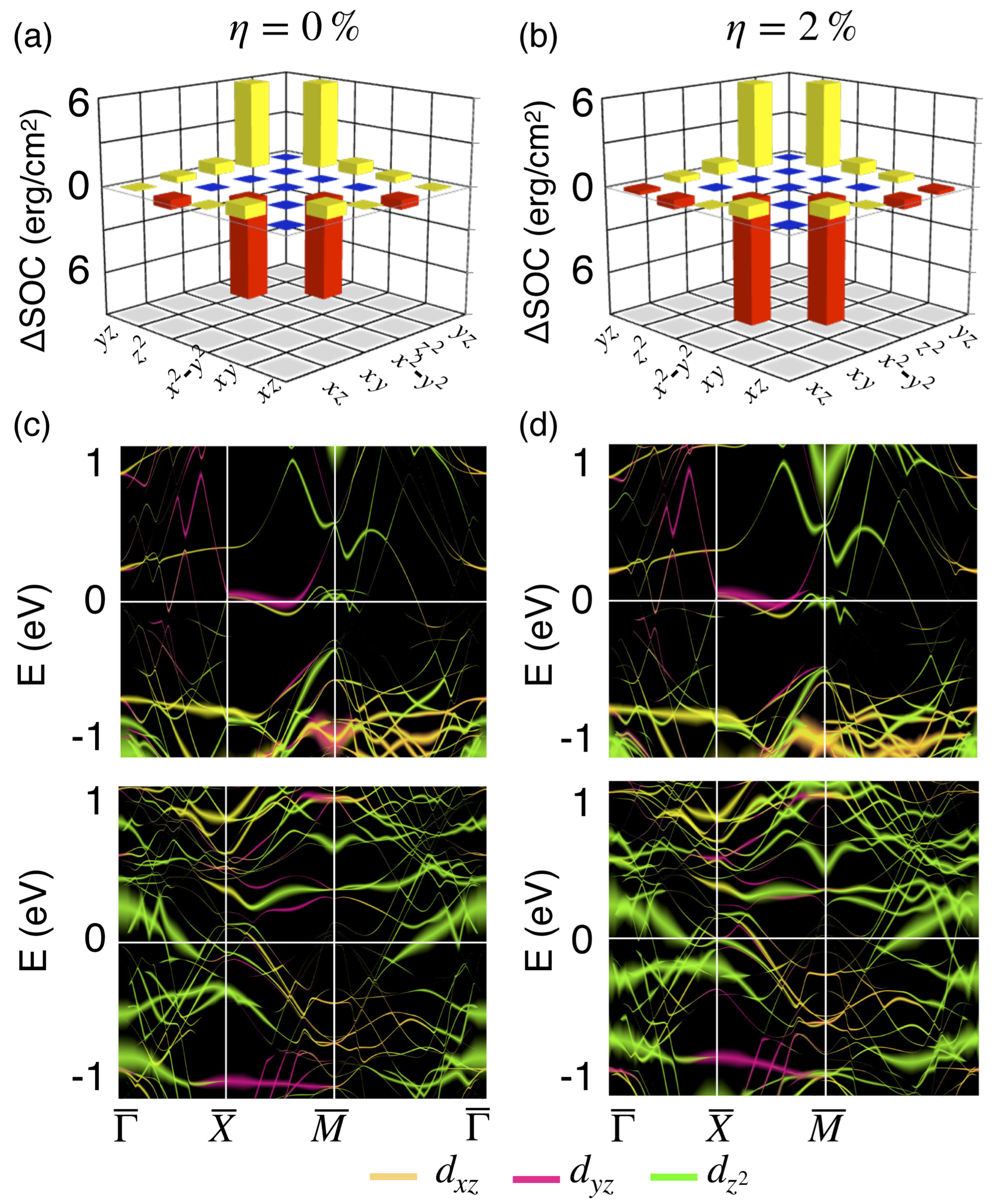}
\caption{\label{fig05} 
Interfacial Pt {\em d}-orbital projected difference in SOC energies ($\Delta$SOC) between in-plane and perpendicular magnetization for polarization $P_+$ at (a) $\eta$ = 0$\%$ and (b) $\eta$ = 2$\%$, where yellow, red, and blue bars represent $\Delta$SOC $>$ 0, $<$ 0, and = 0, respectively.
Orbital-resolved band structure for the majority spin (upper panel) and the minority spin (lower panel) at (c) $\eta$ = 0$\%$ and (d) $\eta$ = 2$\%$.
Bands with yellow, magenta, and green denote $xz$, $yz$, and $z^2$ orbitals, respectively.
The size of the symbols represents the weight of {\em d}-orbitals.  
}
\end{center}
\end{figure}
To elucidate how strain modifies the electronic structure of the interfacial Pt atom, 
the difference in SOC energies ($\Delta$SOC) are compared. 
$\Delta$SOC for $P_+$ at 0$\%$ and 2$\%$ strain are shown in Fig.~\ref{fig05}(a) and (b), respectively. 
Here, $\Delta$SOC denotes the {\em d}-orbital-resolved difference in spin-orbit coupling energy between in-plane and perpendicular magnetization.
To further understand the orbital origin of this variation, Fig.~\ref{fig05}(c) and (d) shows the $d_{xz}$, $d_{yz}$ and $d_{z^2}$ band structure of the interfacial Pt atom along high-symmetry lines in the two-dimensional Brillouin zone for majority and minority spins. 
Contributions from other $d$-orbitals are negligible and are shown in Fig. S3 in the Supplementary Material \cite{BAND}

We employ second-order perturbation theory\cite{wang1993first} to determine $E_{MA}$ via SOC between the occupied and unoccupied states:   
\begin{eqnarray}
\label{eq:EMA}
  E_{\it{MA}}^{\sigma\sigma'}  =  \xi^2\sum_{o,u}\frac{|\langle o^\sigma|\hat{\ell_z}|u^{\sigma'}\rangle|^2-|\langle o^\sigma|\hat{\ell_x}|u^{\sigma'}\rangle|^2}{E_u^{\sigma'} -E_o^\sigma},
\end{eqnarray}
where $\xi$ is the SOC constant, the radial integral of the product of the SOC amplitude $\xi(r)$; $o^\sigma(u^{\sigma'}$) and $E_o^\sigma$($E_u^{\sigma'}$) are the eigenstates and eigenvalues of occupied(unoccupied) states, respectively;
$\hat{\ell_z}$($\hat{\ell_x}$) is the z (x) component of the orbital angular momentum operator.
This method has been successfully adopted in various {\em ab initio} $E_{MA}$ explanations\cite{odkhuu13:prb,odkhuu16:jmmm,odkhuu2018thickness,kim2024strain,ho2022first,qurat2020voltage,daalderop1994magnetic}.

Summing all features in Fig.~\ref{fig05}(a,c) and Fig.~\ref{fig05}(b,d), large $E_{MA}>$ 0 at zero strain is attributed to 
$\langle{yz,xz\uparrow|\hat{L_x}|z^2}\downarrow\rangle$ element along $\overline {XM}$.
These bands remain unaffected under strain.
However, at zero strain, $\langle{x^2-y^2\downarrow|\hat{L_x}|xy}\uparrow\rangle$ along $\overline {M\Gamma}$ yields $E_{MA}<$ 0, and its magnitude increases under strain.
This enhancement in in-plane $E_{MA}$ is evident from the shifting of unoccupied minority $xy$ orbital towards $E_F$ as shown in Fig.~S3 in the Supplementary Material.
As indicated by Eq.~(\ref{eq:EMA}), $E_{MA}$ increases as $(E_u^{\sigma'}-E_o^{\sigma})$ decreases.
All other matrix element contributions almost remain nearly unchanged under strain except for the coupling between $xz$ and $yz$ orbitals.
At $\eta$ = 0$\%$, $\langle{yz\downarrow|\hat{L_z}|xz}\downarrow\rangle$ around the $\overline{X}$ contributes $E_{MA}>$ 0.
Under +2$\%$ strain, the unoccupied $d_{yz}$ band in the minority spin channel at the $\overline{X}$ becomes prominent, contributing $E_{MA}<$ 0 through the $\langle{xz\downarrow|\hat{L_x}|yz}\downarrow\rangle$.
We conclude that strain-induced rearrangement of the interfacial Pt-{\em d} orbital SOC, due to charge redistribution and orbital hybridization, is responsible for the magnetization switching.

\section{Conclusions}
\label{sec:conclusions}

In summary, we presented first-principles study of voltage-controlled magnetization in FePt/BaTiO$_3$ heterostructure. We show that ferroelectric polarization reversal in BaTiO$_3$ induces interfacial orbital reconstruction which induces a significant magnetoelectric response. Our results reveal the strain-mediated switching of magnetic anisotropy, where a competition between interfacial $K_2^i$ and $b_1$ determines magnetization. This competition is governed by strain-induced modifications to spin-orbit coupling of interfacial Pt-$d$ orbitals. The predicted magnetization switching with ferroelectric polarization at 2\% tensile strain, coupled with the large magnetoelectric coefficient. These insights not only clarify the microscopic origin of magnetoelectric coupling in FePt/BTO but also establishes a pathway towards achieving voltage-controlled spintronics through strain-engineering.

\begin{acknowledgments}
  This work was supported by National Research Foundation of Korea (NRF) grant
  (NRF-RS-2024-00449996 and NRF-RS-2024-00451261) and 
  by University Research Fund (URF) of Quaid-i-Azam University, Islamabad (URF-QAU, 2022-2023).  
  We are also grateful for supercomputing Center with supercomputing resources (KSC-2025-CRE-0016) and computational re-  source of the UNIST Supercomputing center.
\end{acknowledgments}
%
\end{document}